\documentclass[preprint,showpacs]{revtex4}
\usepackage{graphics,epsfig}
\usepackage{graphicx}
\usepackage{dcolumn}
\usepackage{amsmath}
\newcommand{\be}{\begin{equation}}
\newcommand{\ee}{\end{equation}}
\newcommand{\beq}{\begin{equation}}
\newcommand{\eeq}{\end{equation}}
\newcommand{\bea}{\begin{eqnarray}}
\newcommand{\eea}{\end{eqnarray}}
\newcommand{\nn}{\nonumber}
\begin{document}
\title{Bardeen-like regular black holes in $ 5D $ Einstein-Gauss-Bonnet gravity}

\author{Dharm Veer Singh}
\email{veerdsingh@gmail.com}
\affiliation{Centre for Theoretical Physics,
Jamia Millia Islamia, New Delhi 110025,
India}

\author{Sushant G. Ghosh}
\email{sgghosh@gmail.com,sghosh2@jmi.ac.in}
\affiliation{Centre for Theoretical Physics,
Jamia Millia Islamia, New Delhi 110025, India}
\affiliation{Multidisciplinary Centre for Advanced Research and Studies (MCARS), Jamia Millia Islamia, New Delhi 110025, India,}
\affiliation{Astrophysics and Cosmology Research Unit,
School of Mathematics, Statistics and Computer Science,
University of KwaZulu-Natal, Private Bag X54001,
Durban 4000, South Africa}

\author{Sunil D. Maharaj}
\email{maharaj@ukzn.ac.za}
\affiliation{Astrophysics and Cosmology Research Unit,
School of Mathematics, Statistics and Computer Science,
University of KwaZulu-Natal, Private Bag X54001,
Durban 4000, South Africa}

\begin{abstract}
We find an exact spherically symmetric regular Bardeen-like solutions by considering the coupling between Einstein-Gauss-Bonnet theory and nonlinear electrodynamics (NED) in five-dimensional spacetime.   These solutions, with an additional parameter $g$ apart from the mass $M$, represent black holes with Cauchy and event horizons, extremal black holes with degenerate horizons or no black holes in the absence of the horizons,   and encompasses as a special case  Boulware-Deser black holes which can be recovered in the absence of magnetic charge ($g=0$). Owing to the NED corrected black hole, the thermodynamic quantities have also been modified and we have obtained exact analytical expressions for the thermodynamical quantities such the Hawking temperature $T_+$,  the entropy $S_+$, the specific heat $C_+$, and the Gibbs free energy $F_+$.  The heat capacity diverges at a critical radius $r=r_C$,  where incidentally the temperature has a maximum, and the Hawking-Page transitions even in absence of the cosmological term. The thermal evaporation process leads to eternal remnants for sufficiently small black holes and  evaporates to a  thermodynamic stable extremel black hole remnants with vanishing temperature. The heat capacity becomes positive $C_+ > 0$ for $r_+ < r_C$ allowing black hole to become thermodynamically stable, in addition the smaller black holes are globally stable with positive heat capacity $C_+ > 0$ and negative free energy $F_+<0$ . The entropy $ S $ of a 5D Bardeen black hole is not longer a quarter of the horizon’s area $A$, i.e., $S \neq A/4$
\end{abstract} 

\maketitle

\section{\label{sec:level1}Introduction}
The cosmic censorship conjecture asserts that physically reasonable matter never creates spacetime  singularities which are observable  to a distant observer \cite{rp}. However, the conjecture does not rule out the possibility of regular (singularity-free) black holes.   Sakharov \cite{Sakharov:1966} and Gliner \cite{Gliner:1966} that suggested one of the ways to avoid  singularities  is to  choose  matter with the equation of state $p=-\rho$,  i.e., matter having a de Sitter core. This could provide a proper discrimination at the final stage of gravitational collapse, replacing the future singularity \cite{Gliner:1966}. Motivated by this idea,  Bardeen \cite{Bardeen:1968} proposed one of the first regular black holes having horizons with no singularity, which  results as an exact solution from Einstein equations coupled to nonlinear electrodynamics (NED) \cite{AGB}.   The spherically symmetric Bardeen black hole is a regular spacetime with a de sitter core, whose metric is given by 
\begin{equation}\label{metric}
ds^2=-\left(1-\frac{2mr^2}{(r^2+g^2)^{3/2}}\right)dt^2+\frac{1}{\left(1-\frac{2mr^2}{(r^2+g^2)^{3/2}}\right)}dr^2+r^2(d\theta^2+\sin^2\theta d\phi^2),
\end{equation}
where $m$ and $g$, are respectively mass and magnetic charge. It turns out that $ g^{rr}=0 $ is a coordinate singularity indicating the possibility of horizons, if they exist,  tare positive roots of $g^{rr}=0$ or 
\begin{eqnarray}\label{eh}
1-\frac{2mr^2}{(r^2+g^2)^{3/2}} =1-\left(\frac{m}{g}\right)\frac{2(r/g)^2}{(1+(r/g)^2)^{3/2}}=0,\;\;\;\;
\text{and}\;\;\;\; r\geq0.
\end{eqnarray}
Solving (\ref{eh}) numerically implies a critical value  $\zeta^{*}$ such that (\ref{eh}) has a double root if $\zeta=\zeta^{*}$, two distinct roots if $\zeta<\zeta^{*}$ and no root if $\zeta>\zeta^{*}$, with $\zeta=m/g$ \cite{Ghosh:2015pra}. These cases illustrate, respectively, an extreme black hole with degenerate horizons, a black hole with Cauchy and event horizons, and no black hole.   We address the regularity of  black hole (\ref{metric})  by calculating the the following scalar invariants
\begin{eqnarray}
&&R_{ab}R^{ab}=\frac{6mg^2(4g^2-r^2)}{(r^2+g^2)^{7/2}},\nonumber \\
&&R_{abcd}R^{abcd}=\frac{12m^2}{(r^2+g^2)^{7/2}}\Big[8g^8-4g^6r^2+47g^4r^4-12g^2r^6+48r^8\Big],
\end{eqnarray}
where $R_{ab}$ and $R_{abcd}$, respectively are Ricci and Riemann tensors. These
invariants  are well behaved everywhere including at $r=0$; hence we can infer that the Bardeen black hole  (\ref{metric}) is a regular. The Bardeen black hole is asymptotically flat, it is according to Schwarzchild  for large  $r$, whereas near the origin it behaves as de Sitter, since \[f(r)\approx 1 - \frac{2m}{g}r^2,\;\; r\approx 0^{+},\]  Thus, the Bardeen black hole, unlike the Schwarzchild  black hole, does not result in a singularity but develops a de Sitter region, eventually settling with a regular center \cite{Borde:1994ai}. Subsequently, there has been intense activity in the investigation of regular black holes \cite{ AyonBeato:1998ub,Xiang,fr1}, and more recently on rotating regular black holes \cite{Toshmatov,Ghosh}. 

One of the natural modification of the general relativity to higher dimensions is by supplementing the Einstein-Hilbert action with second order curvature term, which is so called  Einstein-Gauss-Bonnet (EGB) theory encompasses general relativity as special case. It  is free from the ghosts and also equations of motion is no more than second order and also it describes a wide variety of interesting models  and  hence,  received significant attention.  In particular, EGB gravity includes string theory inspired corrections to the Einstein-Hilbert action and admits Einstein's general relativity as a particular case \cite{gross1}.  Boulware and Deser \cite{Boulware85}  gave an exact static spherically symmetric black holes in EGB theory and demonstrated that the only stable case has a Schwarzschild-type spacetime structure with unavoidable central singularity. The solution was also extended for the charged case \cite{Wiltshare88} with similar conclusion.   Several generalizations of the Boulware-Desser solution with matter source have also been obtained \cite{Dadhich:2012cv,ghosh14}, and from viewpoint of gravitational collapse to a black hole \cite{Dadhich:2013bya}.  The exact spherically symmetric, static,  regular black hole for EGB \cite{dharm}.  However, the Bardeen models are still unexplored in EGB theory.  Since the Bardeen black holes is the  first regular black holes and also opened gate for research in the regular black hoes and it is pertinent to consider  the Bardeen-like black hole in EGB theory.   It the purpose of this paper to construct an exact Bardeen-like black holes in $5D$ EGB theory.  We shall not only  discuss the  horizon of obtained solution, but also investigate the thermodynamic properties including the stability of the system.
 
 Motivated by the development of superstring and other field theories, there is renewed interest in models with extra dimensions. Indeed there are several reasons to study higher-dimensional black holes, in particular, the gravity correspondence \cite{jmm} relates the black holes in $D$ dimensions to that of $D-1$ dimensions and the possibility of producing tiny higher-dimensional black holes at colliders in ”brane-world” scenarios \cite{kanti}. Further, the higher-dimensional black hole spacetimes may have several useful mathematical properties \cite{cvetic}.  The five-dimensional ($5D$) spacetime is particularly more important as spacetime results after dimensional reduction \cite{js}. Also,  the statistical calculation of black hole entropy using string theory was first done for certain $D=5 $  black holes \cite{vafa}.  Hence, the study of higher-dimensional black holes is a  worthwhile contribution in developing a theory of quantum gravity.   In particular, Myers and Perry \cite{mayers} have found solutions to Einstein's equations representing black holes  in $D$ dimensions, and this was extended to Einstien-Maxwell configurations by Dianyan \cite{dianyan}. Other examples with Einstein’s equations are the analysis of spherically symmetric perfect fluids in higher dimensions \cite{Dadhich:2003gw}, in $5D$ spacetimes \cite{Ghosh:2006ab}  and the collapse of different fluids \cite{sgg03}.

The paper is organized as follows. We present relevant equations of EGB theory coupled to NED to obtain an exact $5D$  Bardeen-like black holes EGB theory in Sec. II .  The  investigation of structure and location of the black holes horizons  is the subject of Sec. III.   Sec. IV is devoted to detailed investigation of the thermodynamical properties of  $5D$ EGB-Bardeen black holes. The thermal stability and black hole remnant are in  Sec. V, and the concluding remarks are given in Sec. VI. We shall adopt the the signature $(-,+,+,+,+)$ for the metric and use the units $8\pi G=c=1$.  

\section{ Einstein-Gauss-Bonnet Gravity coupled to nonlinear electrodynamics}
The action for a full interacting EGB theory coupled to NED in $5D$ is given by
\begin{equation}
\label{action} 
\begin{split}
S=\int
d^5x\sqrt{-g}\left[R+\alpha\, (R^2-4R_{ab}R^{ab}+R_{ab c d}R^{ab c d})-{\cal{L}}(F)
\right],
\end{split}
\end{equation}
which is an extension of the Einstein Hilbert action and $\alpha \geq 0$ is the GB coupling constant. The matter is described by NED ${\cal L}(F)$ with $F=F_{ab}F^{ab}/4$, $F_{ab}=\partial_aA_b-\partial_b A_a$. In $4D$ the Euler-Gauss-Bonnet term becomes invariant and it does not contribute to the equation of motion. The action (\ref{action}) can be found in  heterotic superstring theory \cite{gross} in its low energy  limit with $\alpha$ regarded as the inverse string tension \cite{gross,ghosh8}. It is easy to show that the action (\ref{action}), on variation \cite{ghosh8, Hendi:2017phi}, leads to the following field equations of motion
\begin{eqnarray}
&&{G}_{a b} +\alpha {H}_{a b} = T_{ab}\equiv2\left[\frac{\partial {{L(F)}}}{\partial F}F_{a c}F_{b}^{c}-g_{a b}{{L(F)}}\right],\label{egb2}\\
&& \nabla_{a}\left(\frac{\partial {{L(F)}}}{\partial F}F^{a b}\right)=0\qquad \text{and} \qquad \nabla_{\mu}(* F^{ab})=0,
 \label{egb3}
\end{eqnarray}
where $G_{ab}$ and $H_{ab}, $ respectively, are the Einstein tensor
\begin{equation}
G_{ab}=R_{ab}-\frac{1}{2}g_{ab}R,
\end{equation}
and the Lanczos tensor
\begin{equation}
{H}_{ab}=2\Bigl[RR_{ab}-2R_{a c}R^{c}_{b}-2R^{c d
}R_{a c bd} +R_{a}^{~c d e}R_{b c d e}\Bigr]-{1\over 2}g_{ab}{L}_{GB},
\end{equation}
 is divergence free tensor. It is remarkable that the field equations (\ref{egb3}) have derivatives of maximum second order and the theory is free from  ghosts \cite{gross,Ghosh:2016ddh}.  The Maxwell field for NED, with suitable modifications in $5D$ spacetime, reads 
\begin{equation}
F_{ab}=2\delta^{\theta}_{[a}\delta^{\phi}_{b]}Z(r,\theta,\phi),
\label{emt1}
\end{equation}
where $Z$ is function of $r, \theta$ and  $\phi$ in $5D$. It turns out that, in $5D$ spacetime, $F_{\theta\phi}, F_{\theta\psi}$ and $F_{\phi\psi}$ are the nonvanishing components. Integrating Eq. (\ref{egb2}) yields
\begin{equation}
F_{ab}=2\delta^{\theta}_{[a}\delta^{\phi}_{b]}g(r) \sin^2\theta\sin\phi.
\label{emt2}
\end{equation}
Eq. (\ref{egb3}) implies $dF=0$, which leads to $g(r)= g$ = constant.
To obtain our regular black hole solution, we suitably modify Lagrangian density \cite{sabir,ads}
\begin{equation}
{{L(F)}}= \frac{3}{s g^2}\left(\frac{(\sqrt{2 g^2 F})}{(1+\sqrt{2g^2F})}\right)^{7/3}.
\label{nonl1}
\end{equation}
where $s$ is a constant which can be chosen appropriately. When NED Lagrangian ${\cal{L}}(F) \approx F$, the one obtains the $5D$ charged black hole solution \cite{Wiltshare88}.
 Interestingly, the  components of $F_{\theta\phi}$ dominates over the other components and hence can be neglected. Hence, we have
\begin{equation}
F_{\theta\phi}=\frac{g}{r}\sin\theta,\qquad \text{and} \qquad  F=\frac{g^4}{2r^6}.
\label{emt3}
\end{equation}
Using Eq. (\ref{egb3}) and (\ref{emt2}), the energy momentum tensor (EMT) for $5D$ spherically symmetric spacetime is given by
\begin{equation}
T^t_t=T^r_r=\rho(r)= \frac{6 g^5}{s(r^3+g^3)^{7/3}}.
\label{emt4}
\end{equation}
Other components can be easily  obtained using the Bianchi identity $T^{ab}_{;b}=0$ \cite{rizzo06}, 
\begin{equation}
\partial_rT^r_r+\frac{1}{2}g^{00}[T^r_r-T^t_t]\partial_rg_{tt}+\frac{1}{2}\sum g^{ii}[T^r_r-T^i_i]\partial_rg_{ii}=0.
\end{equation}
to obtain
\begin{eqnarray}
T^{\theta}_{\theta}=T^{\phi}_{\phi}=T^{\psi}_{\psi}=\rho(r)+\frac{r}{3}\partial_r\rho(r)=-\frac{2g^5}{s}\frac{3g^3-4r^3}{(r^3+g^3)^{10/3}}.
\label{emt5}
\end{eqnarray}
Thus we have determined complete EMT.

\section{Exact Bardeen-like regular black holes}
Here we are interested in a regular solution of the EGB field Eqs. (\ref{egb2})  with EMTs (\ref{emt4}) and (\ref{emt5})  in a $5D$ spacetime and will  investigate its properties.  We are seeking $5D$ spherically symmetric solution, for which the metric has a form \cite{Ghosh}:
\begin{equation}
ds^2=-f(r)dt^2 +\frac{1}{f(r)}dr^2+r^2d\Omega_3^2,
\label{m1}
\end{equation}
where $d\Omega_3^2=d\theta^2+\sin^2\theta \,(d\phi^2+\sin^2\phi \,d\psi^2)$ denotes the metric on the  $3D$  sphere.  The Einstein field  equations (\ref{egb2}) with metric (\ref{m1}) take the form 
\begin{eqnarray}
&&G^t_t=G^r_r=f'(r)-\frac{2}{r}(1-f(r))+\frac{4\alpha}{r^2}(1-f(r))f'(r)=\frac{6 g^5}{s(r^3+g^3)^{7/3}}, \label{egb4}\\
&&G^{\theta}_{\theta}=G^{\phi}_{\phi}=G^{\psi}_{\psi}=f''(r)+\frac{4}{r}f'(r)+\frac{2}{r^2}(1-f(r))\nn\\&&\qquad\qquad\qquad\qquad+\frac{4\alpha}{r^2}\left[f'(r)(1-f(r))+f'(r)^2\right]=-\frac{2g^5}{s}\frac{3g^3-4r^3}{(r^3+g^3)^{10/3}},
\label{eom1}
\end{eqnarray}  
where a ($^\prime$) means a derivative with respect to  coordinate $r$. It turns out that Eq. (\ref{egb4}) admits an exact solution of the form
\begin{equation}
  f(r)=1+\frac{r^2}{4\alpha}\left(1\pm\sqrt{1+\frac{8 M \alpha }{(r^3+g^3)^{4/3}}}\,\right),
\label{eqn:f}
\end{equation}
where $M$ is the  integration constant related to the black hole mass with dimensions (length)$^2$ and $s=g^2/M$. Further, the metric function (\ref{eqn:f}) satisfies the other field equations.   It turns out that $M=0$ corresponds to
\begin{equation}
f_+=1+\frac{r^2}{2\alpha},\qquad \text {and}\qquad f_-=1.
\end{equation}
Thus, for upper sign (+ {\it ve}) the solution are asymptotically de Sitter and for lower sign (-{\it ve}) they are asymptotically flat. Further, we  note that $+ve$ branch of the  solution is unstable, whereas $-ve$ branch is stable and free from ghost \cite{egb}.  In a realistic physical solution, (\ref{eqn:f}) must become the $5D$ Schwarzschild solution when $\alpha, e \to 0$. However when considering the limit $\alpha,e \to 0$, it does not recover the $5D$ Schwarzschild solution, hence we conclude the (+ {\it ve}) branch solution has no physical interest and hence the (- {\it ve}) branch solution (\ref{m1}) will be discussed.   Thus, we have (\ref{m1}) with metric function (\ref{eqn:f}) and the EGB field equations encompasses the  Boulware-Deser solution \cite{Boulware85}, when $g= 0$. For definiteness, we shall call the solution (\ref{m1}) EGB-Bardeen black holes. The negative branch of the solution,  in the weak limit of GB coupling ($\alpha\rightarrow 0$), becomes
\begin{equation}
ds^2=-\left[1-\frac{M r^2}{(r^3+g^3)^{4/3}}\right]\,dt^2+\frac{1}{\left[1-\frac{M r^2}{(r^3+g^3)^{4/3}}\right]}\,dr^2+r^2\,d\Omega_3^2,
\label{eqn.gb}
\end{equation}
which is $5D$ Bardeen-like solution \cite{sabir}, and goes over to $5D$ Schwarzschild-Tangherilini solution once we switch off the charge ($g=0$). The causal structure of the Bardeen-like solution (\ref{m1}) is similar to that Boulware and Desser \cite{Boulware85} black holes, except it has no more the curvature singularity at $r=0$ \cite{myers88}. The solution (\ref{m1}) can also be understood as a black hole of EGB coupled to NED, henceforth we call it  EGB-Bardeen black holes. For the metric (\ref{m1}), it is seen that the curvature invariants are well behaved and regular with 
\begin{eqnarray}
&&\lim_{r\to 0} R =-\frac{5}{\alpha}+\frac{5}{\alpha}\sqrt{1+\frac{8M\alpha}{g^4}},\nn\\
&&\lim_{r\to 0}R_{ab}R^{ab}=\frac{10}{\alpha^2}+\frac{40}{g^4 \alpha }-\frac{10}{\alpha^2}\sqrt{1+\frac{8M\alpha}{g^4}},\nn\\
&&\lim_{r\to 0} R_{abcd}R^{abcd}=\frac{5}{\alpha^2}+\frac{20}{g^4 \alpha}-\frac{5}{\alpha^2}\sqrt{1+\frac{8M\alpha}{g^4}}.
\label{inv}
\end{eqnarray}
Thus, for $M, \alpha\neq 0$, the invariants are finite everywhere including  at the origin.   The weak energy condition states that $T_{ab}t^at^b\geq 0$ for all timelike vectors $t^a$, or the  local energy density must be non-negative. On the other-hand, the dominant energy condition requires that $T_{ab}t^at^b\geq 0$ for a timelike vector $t^a$. Hence for energy condition  to be satisfied we must have $\rho\geq 0$ and $\rho+P_i\geq 0$. In the $5D$ case,  $T^{ab}t_b$ is spacelike if
\begin{eqnarray}
\rho+P_2=\rho+P_3=\rho+P_4=\frac{6M g^3 }{(r^3+g^3)^{7/3}}\geq 0.
\end{eqnarray}
Clearly,  $\rho > P_3 $  and $P_1 = - \rho$ and hence, EGB-Bardeen black holes obey the energy condition. 

 The event horizon is the largest root of  $g^{rr}=f(r_+)=0$, for which we seek a numerical solution. It is possible to find values of  $\alpha\neq 0$ and $g\neq 0$ such  that $f(r)=0$ admits two positive roots ($r_{\pm}$), where $r_+$ corresponds  to the outer event horizon and $r_-$  to the inner Cauchy horizons. 
\begin{center}
\begin{table}[h]
\begin{center}
\begin{tabular}{l|l r l| r l r l r}
\hline
\hline
\multicolumn{1}{c|}{ }&\multicolumn{1}{c}{ }&\multicolumn{1}{c}{$\alpha=0.1$  }&\multicolumn{1}{c|}{ \,\,\,\,\,\, }&\multicolumn{1}{c}{ }&\multicolumn{1}{c}{}&\multicolumn{1}{c}{ $\alpha=0.2$ }&\multicolumn{1}{c}{}\,\,\,\,\,\,\\
\hline
\multicolumn{1}{c|}{ \it{g}} & \multicolumn{1}{c}{ $r_-$ } & \multicolumn{1}{c}{ $r_+$ }& \multicolumn{1}{c|}{$\delta$}&\multicolumn{1}{c}{\it{ g}}& \multicolumn{1}{c}{$r_-$} &\multicolumn{1}{c}{$r_+$} &\multicolumn{1}{c}{$\delta$}   \\
\hline
\,\, $0$\,\,& \,\,---\,\, &\,\,  0.89\,\,& \,\,---\,\,&0&\,\, ---\,\,&\,\,0.776\,\,&\,\,---\,\,
\\
\
\,\, $0.40$\,\,& \,\,0.37\,\, &\,\,  0.81\,\,& \,\,0.44\,\,&0.25&\,\, 0.27\,\,&\,\,0.74\,\,&\,\,0.47\,\,
\\
\
\,\, $0.45$\,\, & \,\,0.45\,\, &\,\,  0.76\,\,& \,\,0.31&0.30&\,\,0.35\,\,&\,\,0.71\,\,&\,\,0.36\,\,
\\
\
\,\,$g_E=0.493$\,\, &  \,\,0.624\,\,  &\,\,0.624\,\,&\,\,0\,\,&$g_E=0.373$&\,\,0.566\,\,&\,\,0.566\,\,&\,\,0\,\,
\\
\
\,\, $0.55$\,\, & \,\,\,\, &\,\,  \text{No roots}\,\,& &0.45&\,\,\,\,&\,\,\text{No roots}\,\,&\,\,\,\,
\\
\hline
 \hline
\end{tabular}
\end{center}
\caption{Inner and outer horizons and $\delta=r_+-r_-$ for different values of parameter $g$.}
\label{tab:temp1}
\end{table}
\end{center}

\begin{figure*}
\begin{tabular}{c c c c}
\includegraphics[width=.55\linewidth]{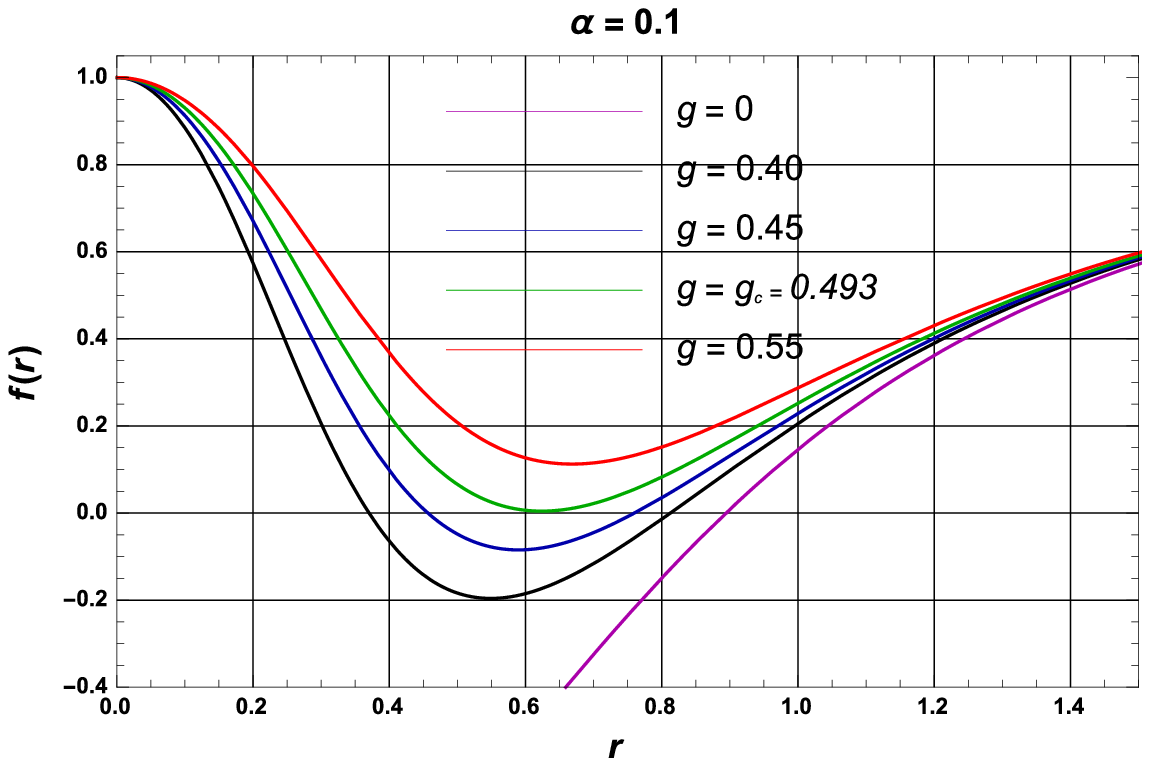}
\includegraphics[width=.55\linewidth]{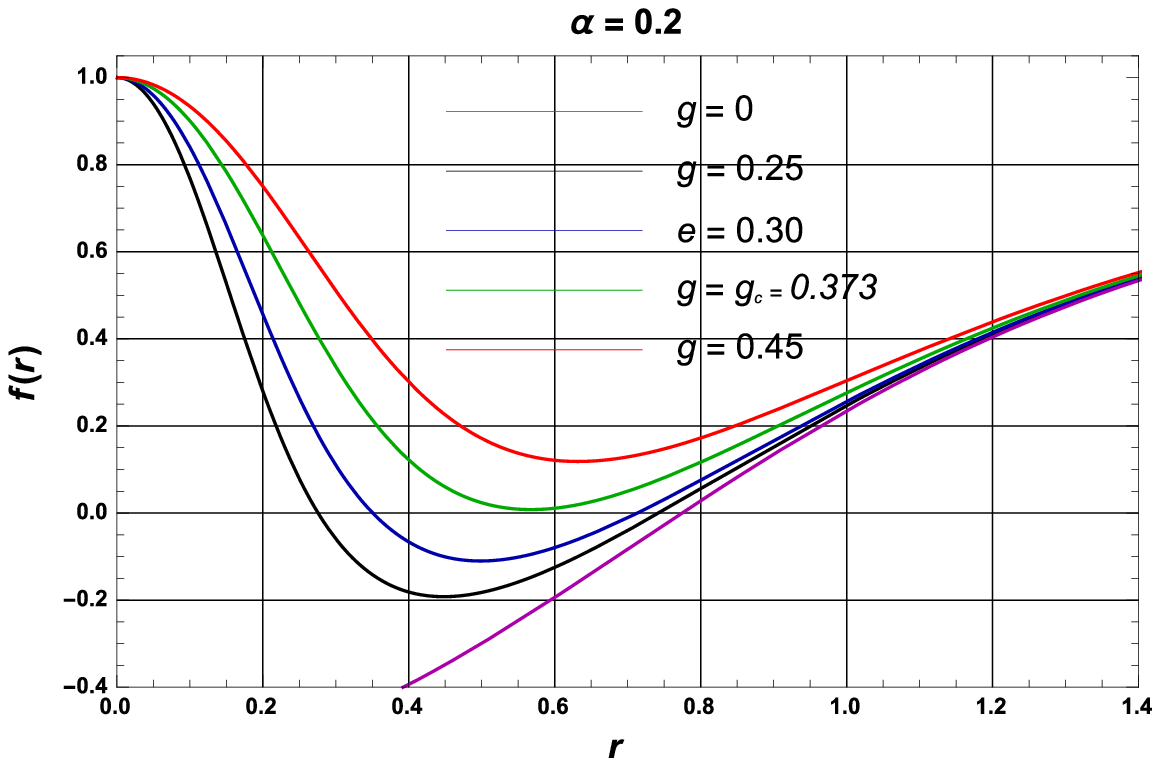}
\end{tabular}
\caption{Metric function $f(r)$  as a function of $r$ for  different values of parameters  $g$ and $\alpha=0.1$ (left) and  $\alpha=0.2$  (right). The critical value of parameter  $g=g_E=0.493$ for $\alpha =0.1$ and $g=g_E=0.373$ for $\alpha=0.2$ respectively. }
\label{fig:1}
\end{figure*}

We have plotted the horizon of the black hole in Fig. \ref{fig:1} and also tabulated some numerical values in Table \ref{tab:temp1}. Fig. \ref{fig:1} suggests,  for a given fixed value of $\alpha$ and $M$, there exists a critical value   $g=g_{E}$ and critical radius $r_E$. Such that $f(r_E)=0,$ which corresponds to the extremal EGB-Bardeen black hole, where two horizons shrink to one, i.e., $r_{\pm}=r_E$ such that $f(r_E)=f'(r_E)=0$. 

\section{Black Hole Thermodynamics}
The thermodynamics  of black holes  is useful to throw light on the  quantum properties of the gravitational field.  The thermodynamics of AdS black holes have been of great interest due to the  existence of a phase transition  in  AdS black holes \cite{hp}.    Here, we  are interested in the  thermodynamical properties of  the $5D$  EGB-Bardeen  black holes, which is characterized by mass $M$,  Gauss-Bonnet coupling $\alpha$ and  parameter $g$.  The black hole mass $M$  using Eq. (\ref{eqn:f}),  in terms of the horizon radius $r_+$, reads
\begin{equation}
M_+=r_+^2\left(1+\frac{2\alpha}{r_+^2}\right)\left(1+\frac{g^3}{r_+^3}\right)^{4/3},
\label{eqM}
\end{equation}
\noindent   when $g=0,$ we have  $ M_+=r_+^2+2\alpha$ the EGB-Bardeen black hole mass\cite{cai02,ghosh8,ghosh14}  and further  for $\alpha=0$, we have $M_+=r_+^2$ \cite{ghosh8}  the  $5 D$ Schwarzschild-Tangherlini black hole.  The surface gravity is
\begin{equation} 
\kappa=\frac{1}{2\pi}\left(-\frac{1}{2}\nabla_{\mu}\xi_{\nu}\nabla^{\mu}\xi^{\nu}\right)^{1/2},
\label{temp21}
\end{equation}
and $\xi^{\mu}=\partial/\partial t$ is a Killing vector. 
Now we are ready to analyze the thermodynamic quantities of $5D$ EGB-Bardeen black hole.
The Hawking temperature  $5D$  EGB-Bardeen  black hole is  $T=\kappa/2\pi$ with  $\kappa$ the surface gravity. The EGB-Bardeen black hole temperature, using (\ref{m1}), (\ref{eqn:f}) and (\ref{temp21}), reads
\begin{equation}
T_+= \frac{f'(r_+)}{4\pi}=\frac{1}{2\pi r_+}\left[\frac{r_+^2-\frac{g^3}{r_+^3}\left(r_+^2+{4\alpha}\right)}{(1+\frac{g^3}{r_+^3})(r_+^2+4\alpha)}\right],
\label{eqT}
\end{equation}
\begin{figure*}[ht]
\begin{tabular}{c c c c}
\includegraphics[width=0.55\linewidth]{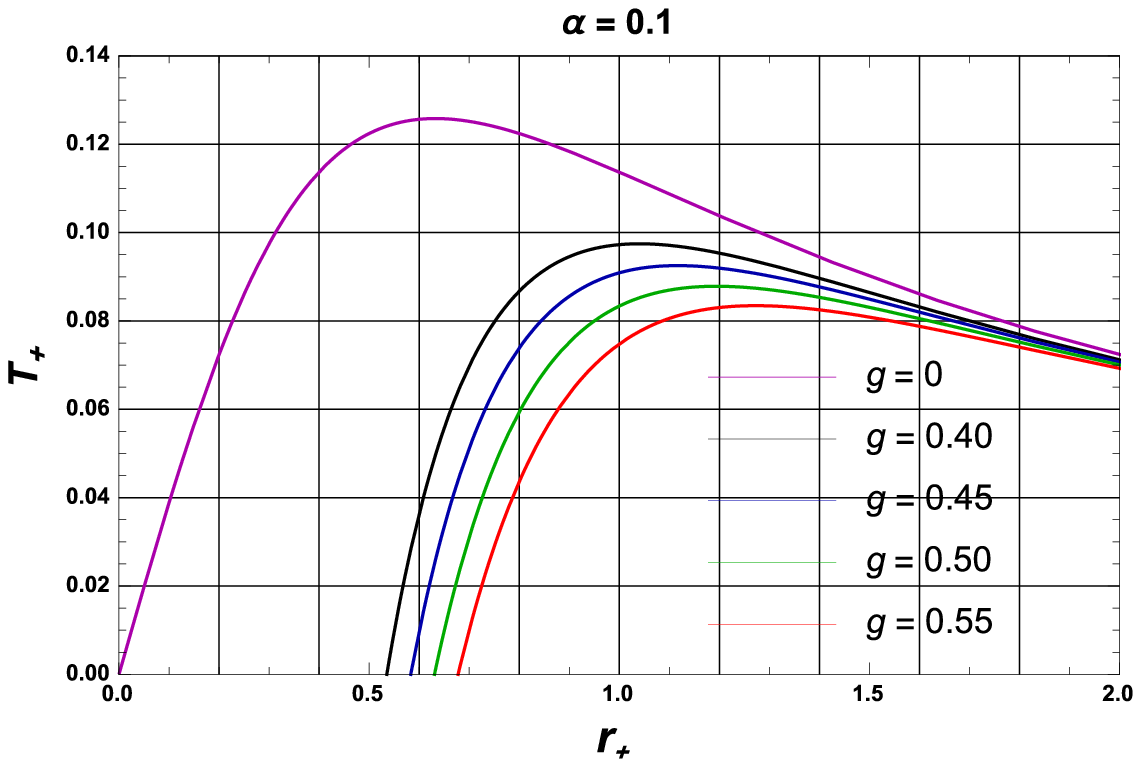}
\includegraphics[width=0.55\linewidth]{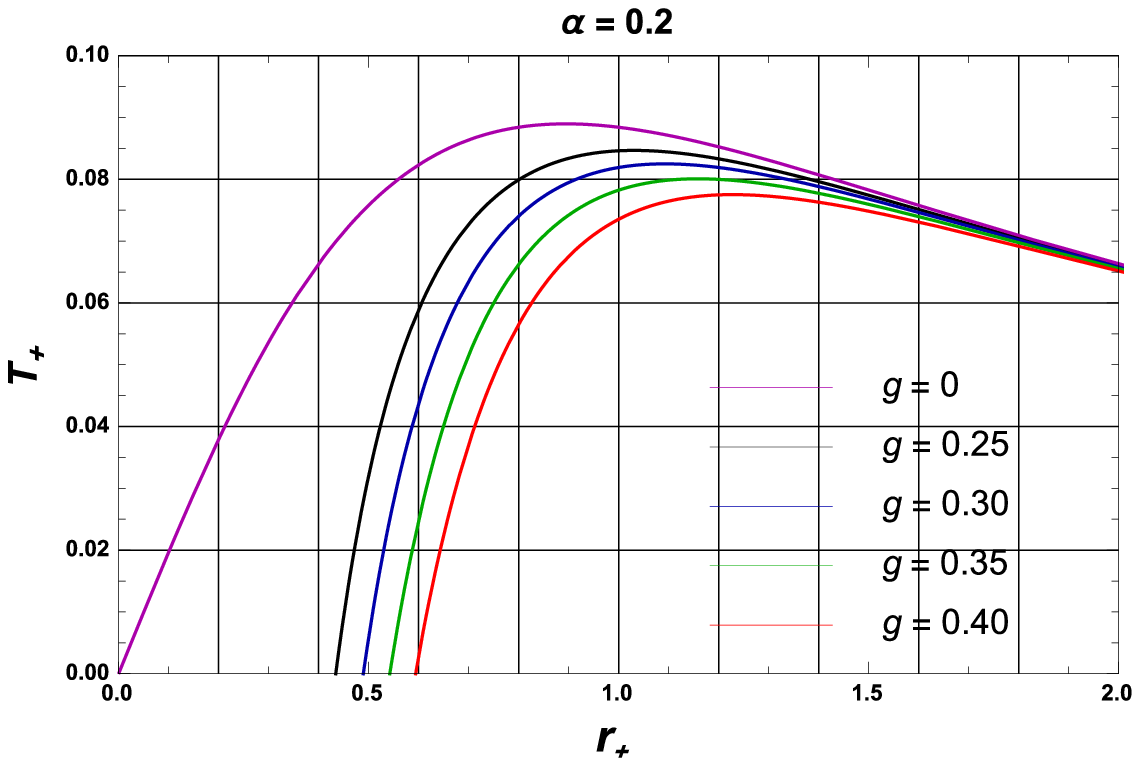}
\end{tabular}
\caption{Hawking temperature $T_+$  as a function of horizon radius $r_+$  for  different values of parameter $g$ .}
\label{fig:th1}
\end{figure*}
Obviously positive temperature $T_+>0$ requires $g^3>r_+^5/r_+^2+4\alpha$.  The temperature vanishes if the black hole is extremal  degenerate horizons, i.e., when $r_+=r_-$. The Hawking temperature has a peak (cf. Fig. \ref{fig:th1} and Table \ref{tab:temp}) that shifts to right and decreases with increasing $g$. The temperature (\ref{eqT}), in the absence of the parameter ($g=0$), becomes 
\begin{equation}
T_+=\frac{1}{2\pi }\left(\frac{r_+}{r_+^2+4\alpha}\right),
\end{equation}
It is the temperature of the  EGB  black hole \cite{cai02,ghosh8,ghosh14}. Further, in the limit $\alpha \to 0$, the temperature (\ref{eqT}) reduces to the case of $5D$  Bardeen black hole \cite{sabir}
\begin{equation}
T_+= \frac{1}{2\pi r_+}\left(\frac{1-\frac{g^3}{r_+^3}}{1+\frac{g^3}{r_+^3}}\right).
\label{eqT1}
\end{equation}
\begin{figure*}[ht]
\begin{tabular}{c c c c}
\includegraphics[width=0.7\linewidth]{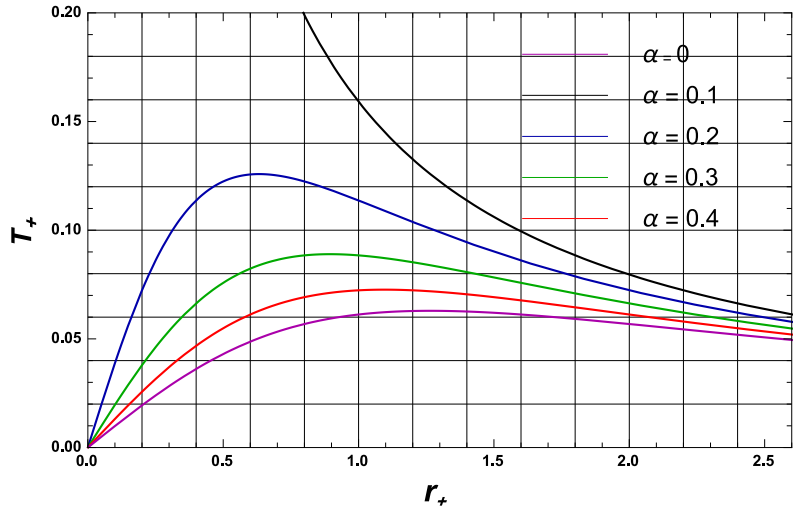}
\end{tabular}
\caption{Temperature $T_+$ as a function horizon radius $r_+$ for $5D$  EGB-Bardeen black holes for  different values of  $\alpha$.}
\label{fig:th2}
\end{figure*}
 Figs. \ref{fig:th1} and \ref{fig:th2} show that the temperature grows to a maximum $T_{max}$  then drops to zero temperature.   The Hawking temperature  of the $5D$  EGB-Bardeen  black hole has maximum at the critical radius shown in Table \ref{tab:temp}. When we increase ithe values of $g$ and $\alpha$, the local maximum of the Hawking temperature decreases  (cf. Fig. \ref{fig:th1}). Further, the temperature diverges when the horizon radius shrinks to zero for $5D$ Schwarzschild black hole  (cf. Fig. \ref{fig:th2}).  In turn, the temperature yields $T_+=1/2\pi r_+$ when  $g=0$ for the $5D$ Schwarzschild-Tangherilini black holes (\ref{eqT1}) \cite{ghosh8,ghosh14}. 
\begin{center}
\begin{table}[h]
\begin{center}
\begin{tabular}{l|l r l r l| r l r r r}
\hline
\hline
\multicolumn{1}{c}{ }&\multicolumn{1}{c}{ }&\multicolumn{1}{c}{ }&\multicolumn{1}{c}{$\alpha=0.1$  }&\multicolumn{1}{c}{ \,\,\,\,\,\, }&\multicolumn{1}{c|}{ }&\multicolumn{1}{c}{  }&\multicolumn{1}{c}{ }&\multicolumn{1}{c}{$\alpha=0.2$}\,\,\,\,\,\,\\
\hline
\multicolumn{1}{c|}{ \it{g} } &\multicolumn{1}{c}{ 0 } &\multicolumn{1}{c}{ 0.40 } & \multicolumn{1}{c}{ 0.45 }& \multicolumn{1}{c}{0.50}& \multicolumn{1}{c|}{0.55} &\multicolumn{1}{c}{ 0}&\multicolumn{1}{c}{0.25} &\multicolumn{1}{c}{ 0.30}   & \multicolumn{1}{c}{0.35}& \multicolumn{1}{c}{0.40} \\
\hline
\,\,$r_c^T$\,\, &\,\,0.63\,\,&  \,\, 1\,\, &\,\,1.036\,\, &  \,\,1.11\,\, &\,\,1.12\,\,&\,\,0.897\,\,&\,\,0.99\,\,&\,\,1.00\,\,&\,\,1.14\,\,&\,\,1.20\,\,
 \\
\,\,$T_+^{Max}$\,\,&\,\,0.125\,\,&\,\,  0.097\,\, & \,\,0.091\,\,& \,\,  0.087\,\, &\,\,  0.083\,\,&\,\,0.088\,\,& \,\,0.084\,\,&\,\, 0.081\,\,&\,\,0.079\,\,&\,\,0.076\,\,
\\
\hline
 \hline
\end{tabular}
\end{center}
\caption{Maximum Hawking temperature ($T_+^{Max}$) at critical radius ($r_c^{T}$) for different  $g$.}
\label{tab:temp}
\end{table}
\end{center}
The entropy  associated  with a black hole,  can be obtained by the first law of thermodynamics \cite{ghosh8,ghosh14}
\begin{equation}
dM_+=T_+\,dS_++\phi\, dg,
\end{equation}
where $\phi$ is the potential for the constant parameter $g$ \cite{Maluf:2018lyu}.  Thus, the entropy of   $5D$ EGB-Bardeen black hole 
\begin{eqnarray}
S_+=\int\,\frac{1}{T_+}\frac{\partial M_+}{\partial r_+}dr_+,
\label{eqS}
\end{eqnarray}
and substituting (\ref{eqM}) and (\ref{eqT}) into (\ref{eqS}), an exact expression for the entropy of the $5D$ EGB Bardeen  black hole reads
\begin{eqnarray}
S_+&=&\frac{4\pi r_+}{3}\Bigg[\left(1+\frac{g^3}{r_+^3}\right)^{1/3}\left(r_+^2+12\alpha-\frac{g^3}{r_+^3}(3r_+^2+4\alpha)\right)\nn\\&&\qquad\qquad-\frac{8\alpha g^3}{r_+^5}2F_1\left[\frac{2}{3},\frac{2}{3},\frac{5}{3},-\frac{g^3}{r_+^3}\right]+\frac{2g}{r_+}2F_1\left[\frac{2}{3},\frac{2}{3},\frac{5}{3},-\frac{r_+^3}{g^3}\right]\Bigg].
\label{entropy}
\end{eqnarray}
The  expression of entropy Eq. (\ref{entropy}) is modified due  parameters $g$ and  $\alpha$ with the  identification between entropy and area is no longer valid for $5D$ EGB-Bardeen black holes.

In the absence of   NED  $(g=0)$, the entropy becomes
\begin{eqnarray}
S_+=\frac{4\pi r_+^3}{3} +16\alpha r_+,
\label{entropy1}
\end{eqnarray}
 the entropy of the EGB-Bardeen black hole \cite{cai02,ghosh8,ghosh14}.  Further, when $\alpha \to 0$, the entropy is exactly same as the $5D$ Bardeen black hole \cite{sabir}.  When both $g=0,\alpha\to 0$, one sees  $S_+=4\pi r_+^3/3$ as the entropy of the $5D$ Schwarzschild-Tangherilini black hole \cite{ghosh8} and area law is valid. 
\section{Thermodynamical stability and black hole remnant}
 It is of considerable interest to study the stability of a given static field configuration, and focus on the local stability or thermal stability. A black hole is stable/unstable when heat capacity $C_+$  is positive/negative \cite{Antoniou:2017hxj}.  The  thermodynamical stability of a black hole is performed by studying the behavior of  its heat capacity ($C_+$), as the local stability of black hole is related to its heat capacity. The  heat capacity  of a black hole is given by \cite{ ghosh8,ghosh14}
\begin{eqnarray}
C_+=\frac{\partial M_+}{\partial T_+}=\Big(\frac{\partial M_+}{\partial r_+}\Big)\Big(\frac{\partial r_+}{\partial T_+}\Big).
\label{eqC}
\end{eqnarray}
Substituting (\ref{eqM}) and (\ref{eqT}) into (\ref{eqC}),  we obtain the heat capacity of the $5D$  EGB Bardeen black hole as
\begin{equation}
C_+=-4\pi r_+^3\left[\frac{(r_+^2+4\alpha)^2(r_+^3+g^3)^{\frac{7}{3}}\left(r_+^2-\frac{g^3}{r_+^3}(r_+^2+4\alpha)\right)}{[r_+^3(g^3(48\alpha r_+^5+8\alpha r_+^2 g^3+6r_+^7+r_+^4g^3+64\alpha^2 r_+^3+16\alpha^2g^3)-r_+^8(r_+^2-4\alpha))]}\right].
\label{eqc1}
\end{equation}
\begin{figure*}[ht]
\begin{tabular}{c c c c}
\includegraphics[width=0.50\linewidth]{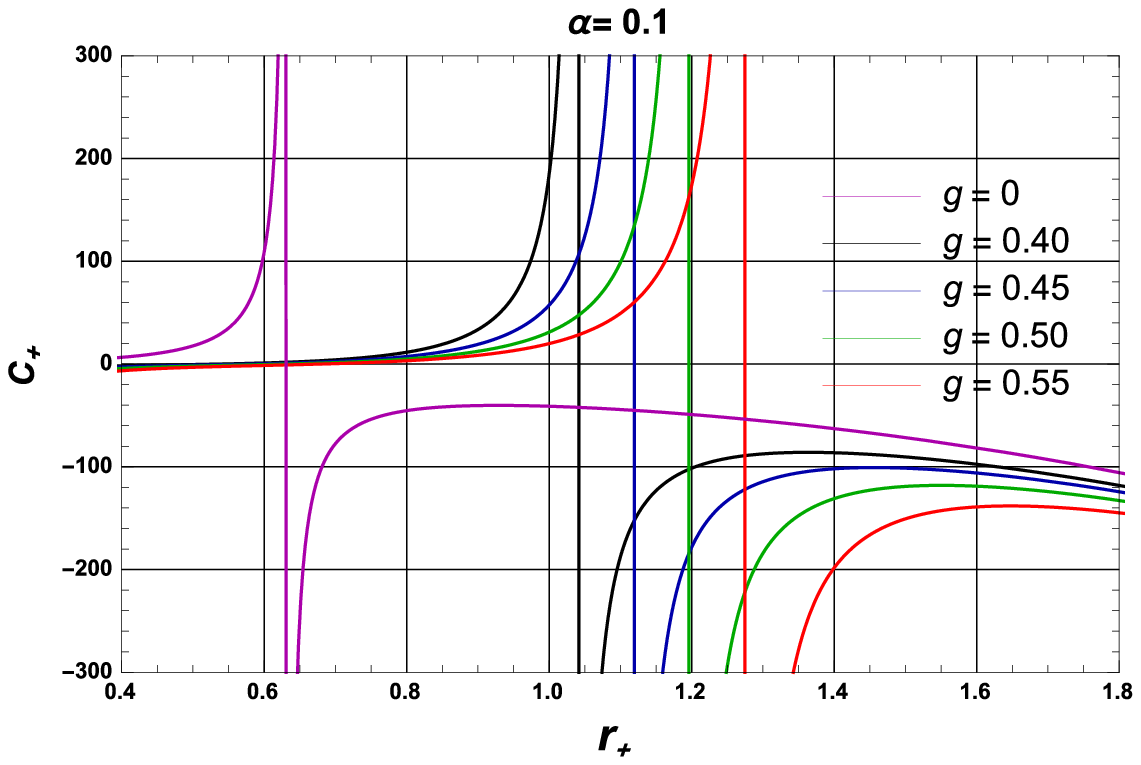}
\includegraphics[width=0.50\linewidth]{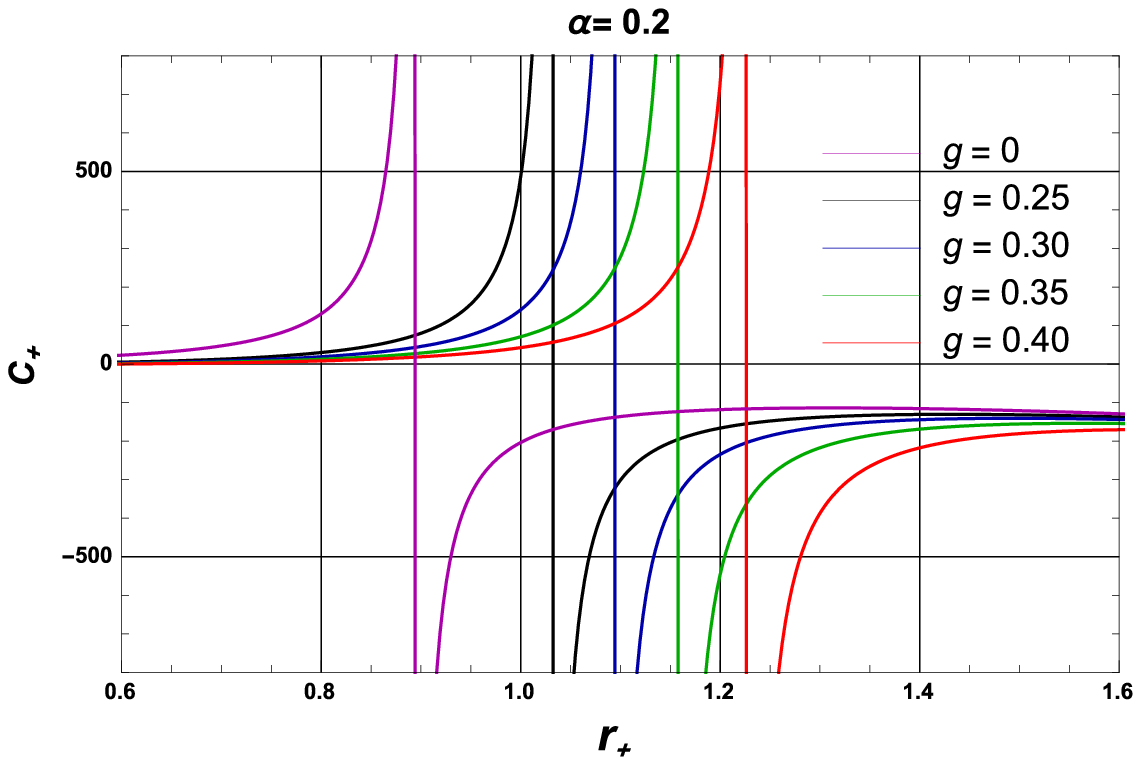}\\
\includegraphics[width=0.50\linewidth]{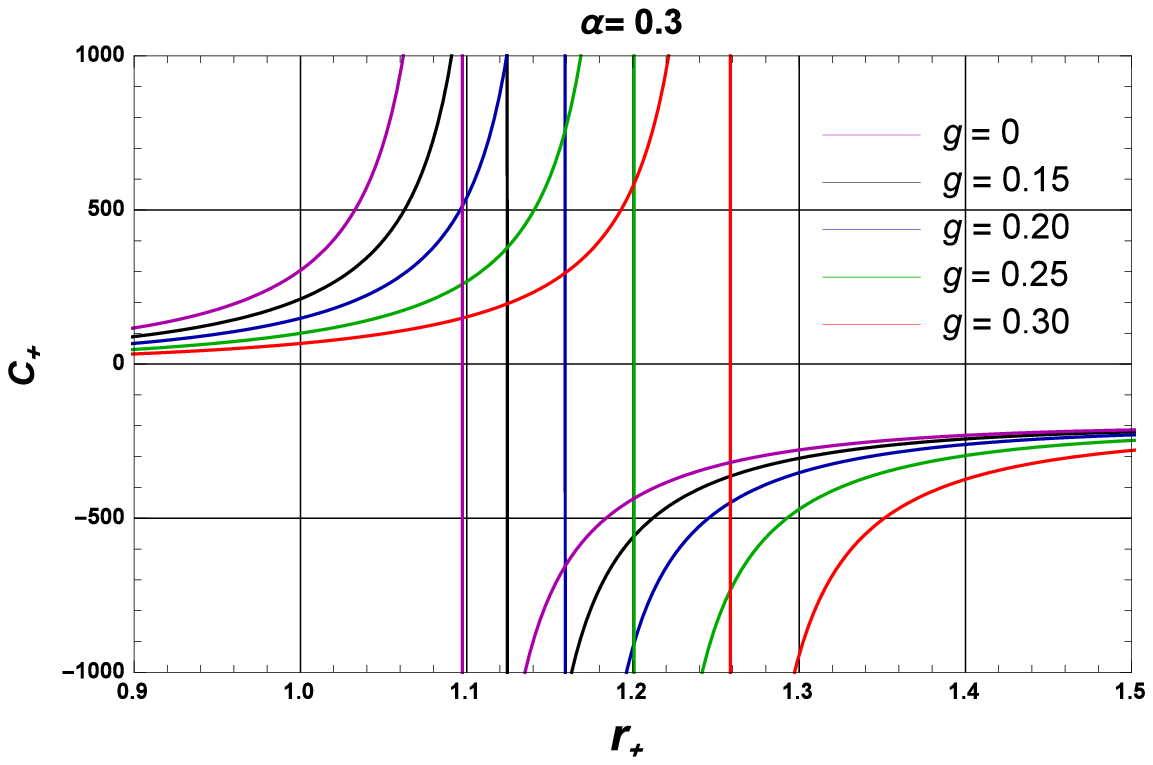}
\includegraphics[width=0.50\linewidth]{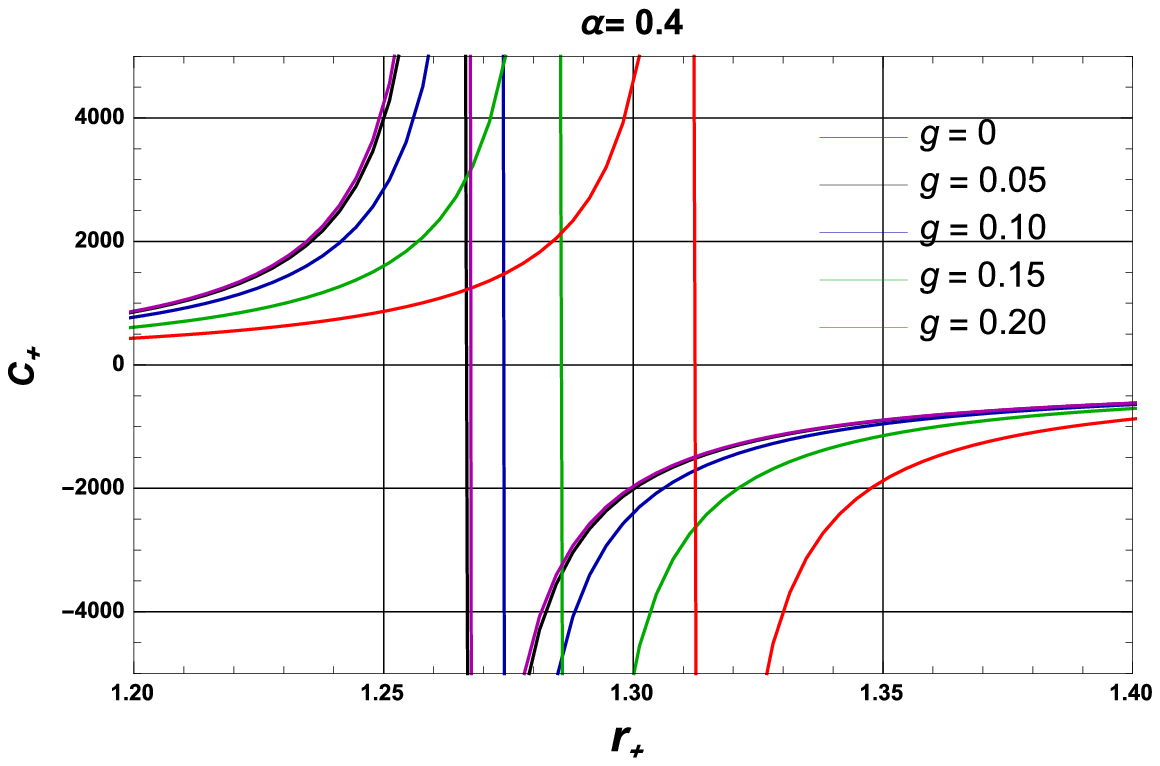}
\end{tabular}
\caption{\label{fig:sh}The specific heat $C_+$ as a function of horizon radius $r_+$  for different values of  parameters  $g$  and $\alpha$.}
\end{figure*}
The heat capacity,  for different values of parameter $g$ and $\alpha$, is depicted  in the Fig. \ref{fig:sh} . We observe that the heat capacity is positive for $r_+<r_C$ suggesting that the  black hole is  thermodynamically stable and is negative heat capacity for $r_+ > r_C$, implying the instability of black holes, with heat capacity  discontinuous  at critical $r_+=r_C$, which implies second order phase transition happens \cite{hp,davis77}. A discontinuity of the heat capacity occur  at $r_+=1.11$, at which the Hawking temperature  maximum value $T_+=0.087$ for $\alpha=0.1$ and $g=0.493$  (Fig. \ref{fig:sh}). The critical radius $r_C$ increases with parameter $\alpha$ (cf. Fig. \ref{fig:sh} and Table \ref{tab:temp}). The EGB-Bardeen black hole has stable region $r_+ < r_C$ and heat capacity diverges at $r_+=r_C$ when the temperature is maximum.  In the limit $g \to 0$ one obtains the heat capacity for the EGB-Bardeen black holes
\begin{equation}
C_+=-4\pi r_+^3\frac{(r_+^2+4\alpha)^2}{r_+^4-4\alpha r_+^2}.
\label{cgb}
\end{equation}
In the limit $\alpha \to0$, the heat capacity (\ref{eqc1}) reduces to
 \begin{equation}\label{cgb1}
 C_+=-4\pi r_+^3\left[\frac{r_+(r_+^3+g^3)^{\frac{7}{3}}\left(r_+^2-\frac{g^3}{r_+}\right)}{r_+^{7}-g^3(6r_+^7+r_+^4g^3)}\right],
 \end{equation}
 which is exactly same as $5D$ Bardeen black hole \cite{sabir}, and (\ref{cgb1}) with $g=0$ corresponds to the heat capacity of the $5D$ Schwarzschild-Tangherilini black hole \cite{ghosh8,ghosh14}.   From Eq. (\ref{cgb}), we observe that $C_+ \rightarrow \infty$ as $r_+^2 \rightarrow 4\alpha $ and also at $ r_+^2=4\alpha $ the heat capacity flips the sign, representing a  transition point (cf. Fig. \ref{fig:sh}). 

One can also calculate the Gibbs free energy to discuss the global stability of the black holes \cite{Herscovich} given by
\begin{equation}
F_+ =M_+-T_+S_+.
\label{freef}
\end{equation}
Substituting the values of Eqs. (\ref{eqM}),  (\ref{eqT}) and  (\ref{entropy}) in Eq. (\ref{freef}), we get
\begin{eqnarray}
F_+=&&\left(r_+^2+{2\alpha}\right)\left(1+\frac{g^3}{r_+^3}\right)^{4/3}-\left(\frac{r_+^5-g^3(r_+^2+4\alpha)}{(4\alpha+r_+^2)(r_+^3+g^3)}\right)\Bigg[\frac{4\pi r_+}{3}\Bigg[\left(1+\frac{g^3}{r_+^3}\right)^{1/3}\nn\\&&\left(r_+^2+12\alpha-\frac{g^3}{r_+^3}(3r_+^2+4\alpha)\right)-\frac{8\alpha g^3}{r_+^5}2F_1\left[\frac{2}{3},\frac{2}{3},\frac{5}{3},-\frac{g^3}{r_+^3}\right]+\frac{2g}{r_+}2F_1\left[\frac{2}{3},\frac{2}{3},\frac{5}{3},-\frac{r_+^3}{g^3}\right]\Bigg]\Bigg].\nn\\
\end{eqnarray}
In the limit $\alpha \to 0$, gives the Gibbs free energy for the $5D$  Bardeen black hole as
\begin{eqnarray}
F_+=&&r_+^2\left(1+\frac{g^3}{r_+^3}\right)^{4/3}-\frac{4\pi r_+}{3}\left(\frac{r_+^3-g^3}{r_+^3+g^3}\right)\Bigg[\left(1+\frac{g^3}{r_+^3}\right)^{1/3}\left(r_+^3-\frac{3g^3}{r_+^3}\right)+\frac{2g}{r_+}2F_1\left[\frac{2}{3},\frac{2}{3},\frac{5}{3},-\frac{r_+^3}{g^3}\right]\Bigg].\nn\\
\end{eqnarray}
\begin{figure*}[ht]
\begin{tabular}{c c c c}
\includegraphics[width=0.55\linewidth]{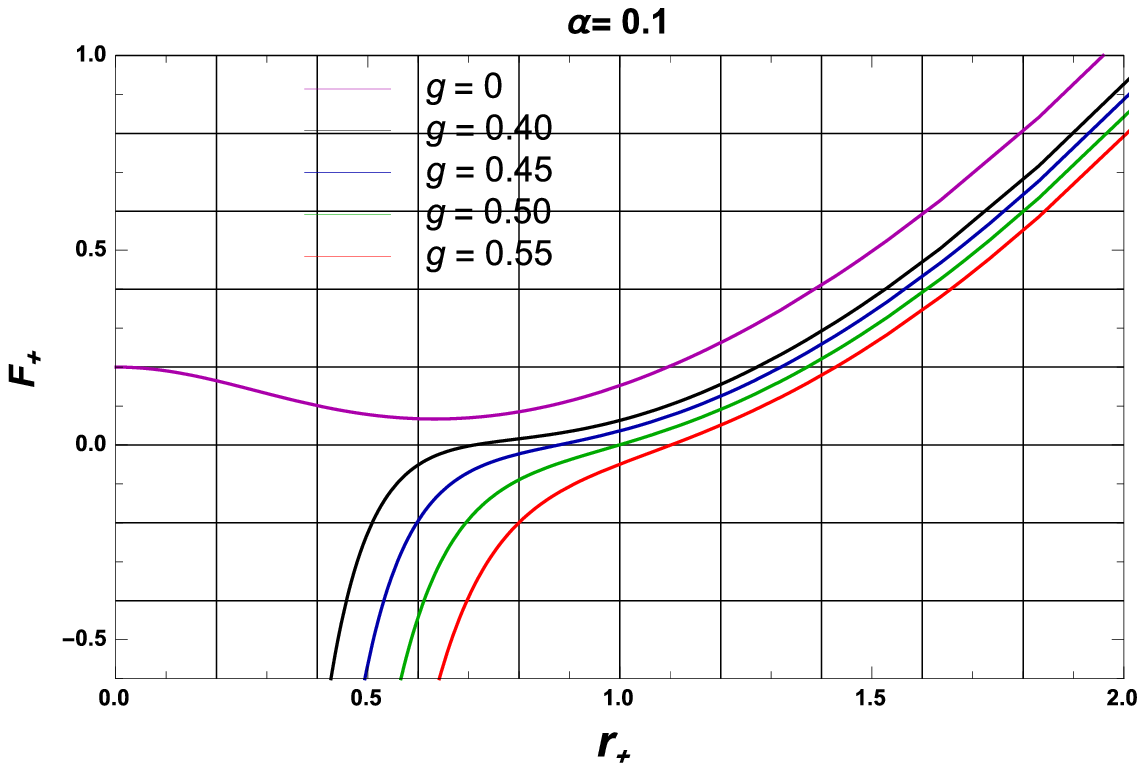}
\includegraphics[width=0.55\linewidth]{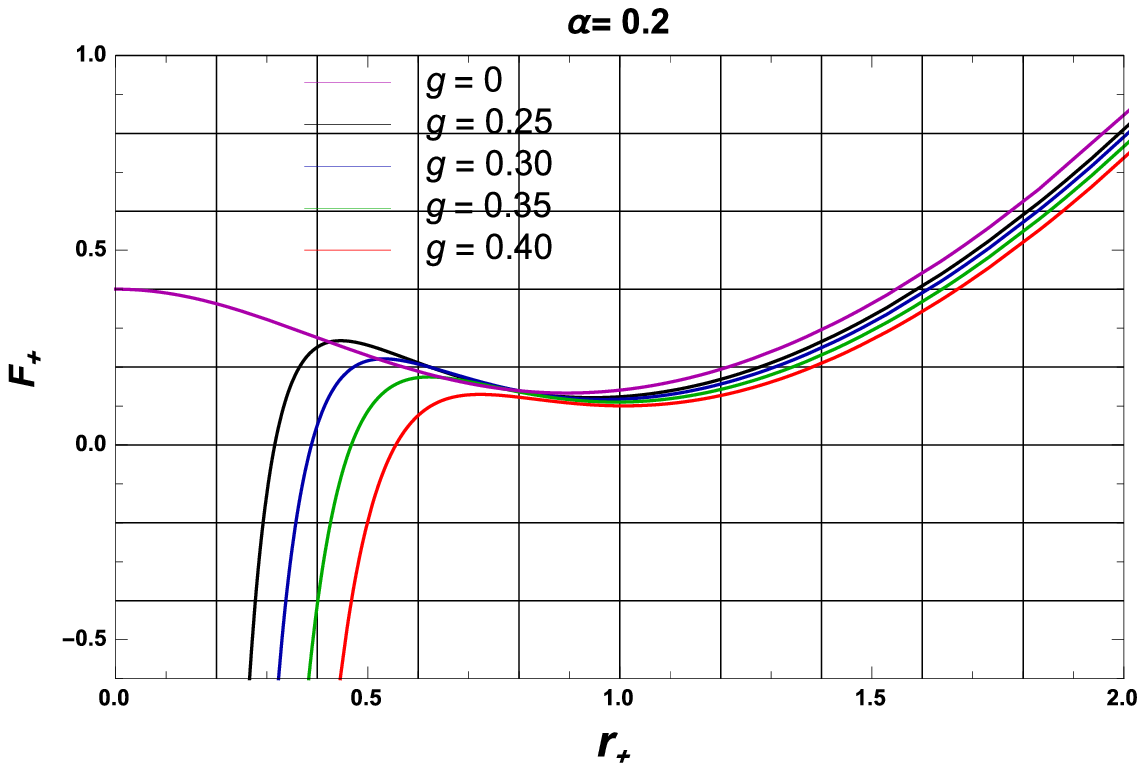}
\end{tabular}
\caption{Free energy $F_+$ as a function of horizon radius $r_+$ for  different values of  parameters $g$  and $\alpha$.}
\label{free1}
\end{figure*}
The Gibbs free energy for various value of parameter $g$ is shown in Fig. \ref{free1} which shows that $F_+<0 $ for smaller $r_+$ where the heat capacity $C_+>0$ , and thus indicating that  EGB-Bardeen black holes are stable for smaller radius  $r_+$.

The black hole remnant is a well-merited apprehension in theoretical astrophysics, which is one of the candidates to resolve the information loss puzzle \cite{jp} and also a source for the dark energy \cite{jh}. As pointed above, The double root $r_-=r_E$ of $f(r)=0$ corresponds to the extremal black hole with degenerate horizon and hence $f'(r_E)=0$.
\begin{figure*}[ht]
\begin{tabular}{c c c c}
\includegraphics[width=0.55\linewidth]{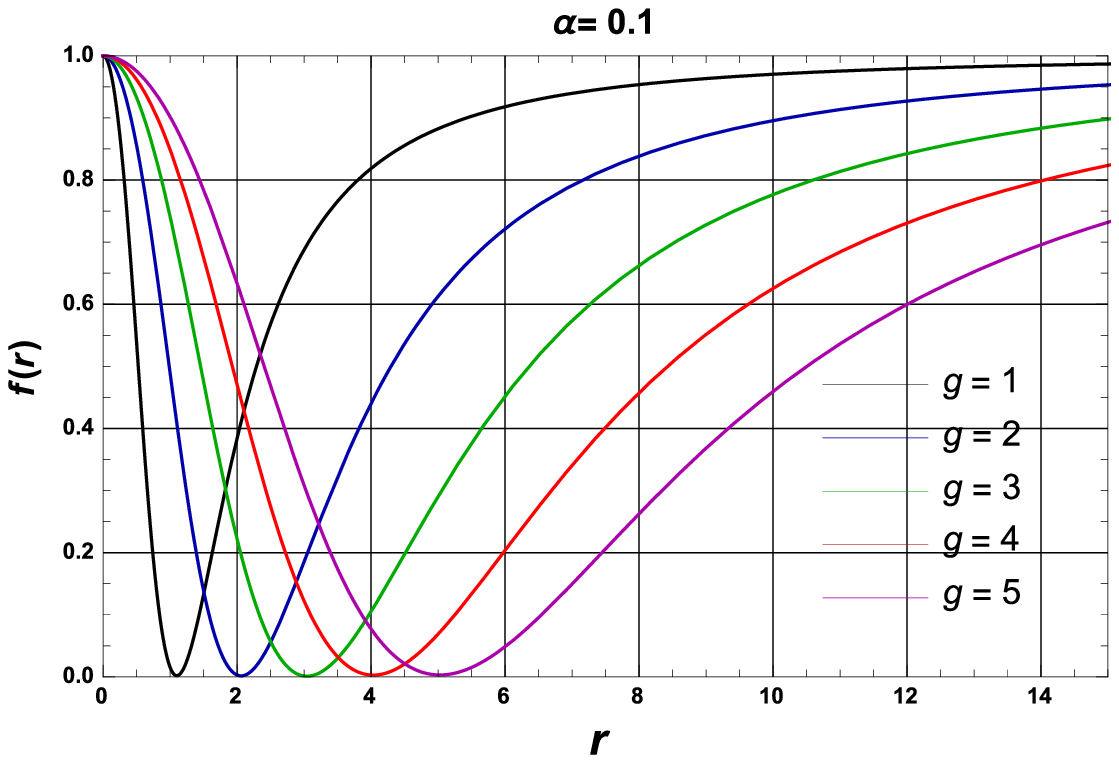}
\includegraphics[width=0.55\linewidth]{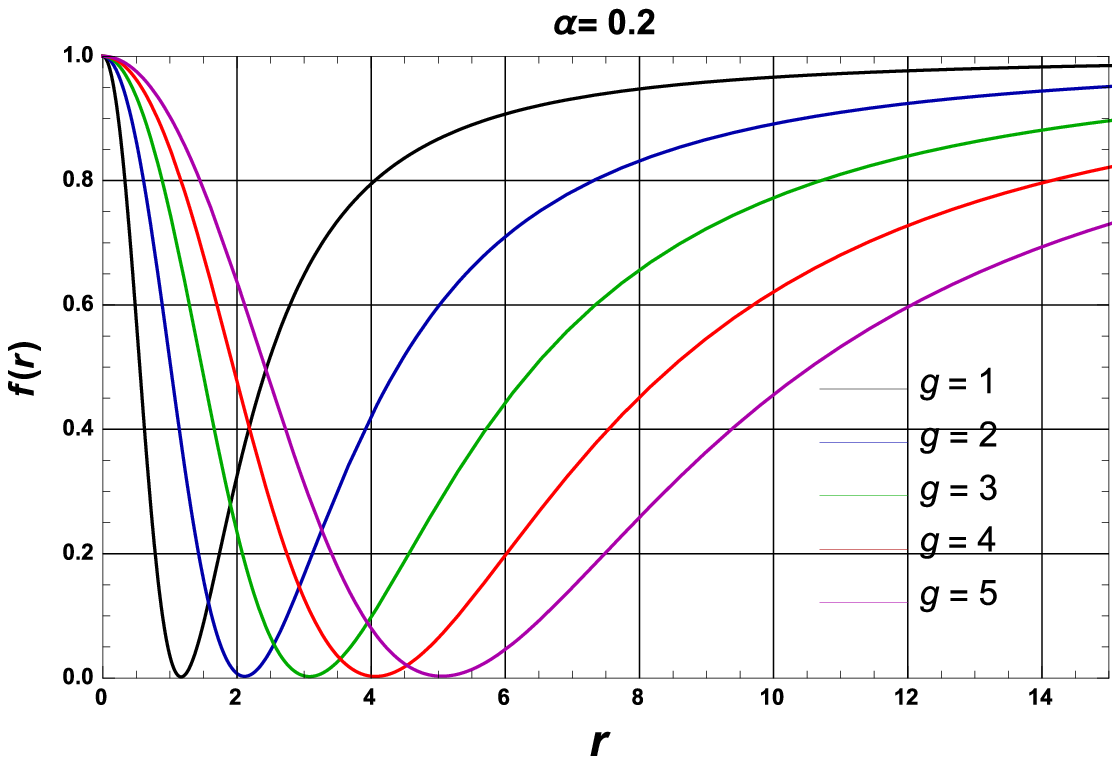}
\end{tabular}
\caption{The plot of metric function $f(r)$ as the function horizon radius $r_+$ for  different values of  parameters $g$  and $\alpha$.}
\label{rem1}
\end{figure*}
\begin{figure*}[ht]
\begin{tabular}{c c c c}
\includegraphics[width=0.55\linewidth]{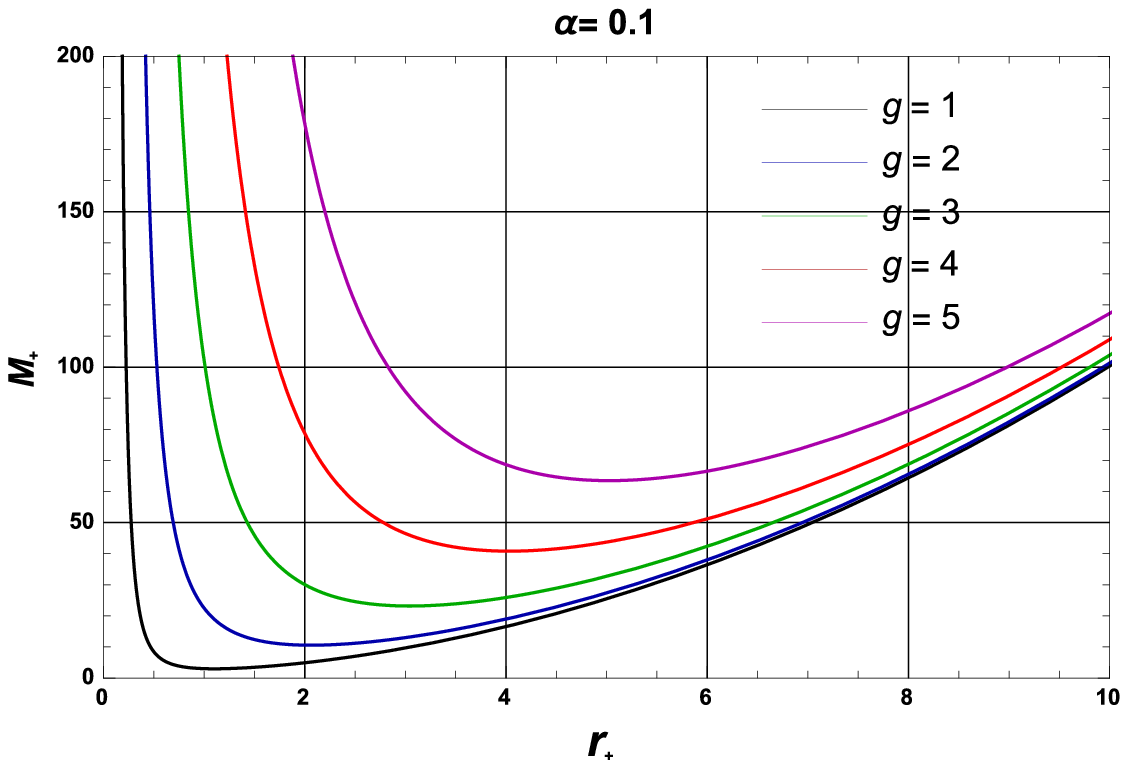}
\includegraphics[width=0.55\linewidth]{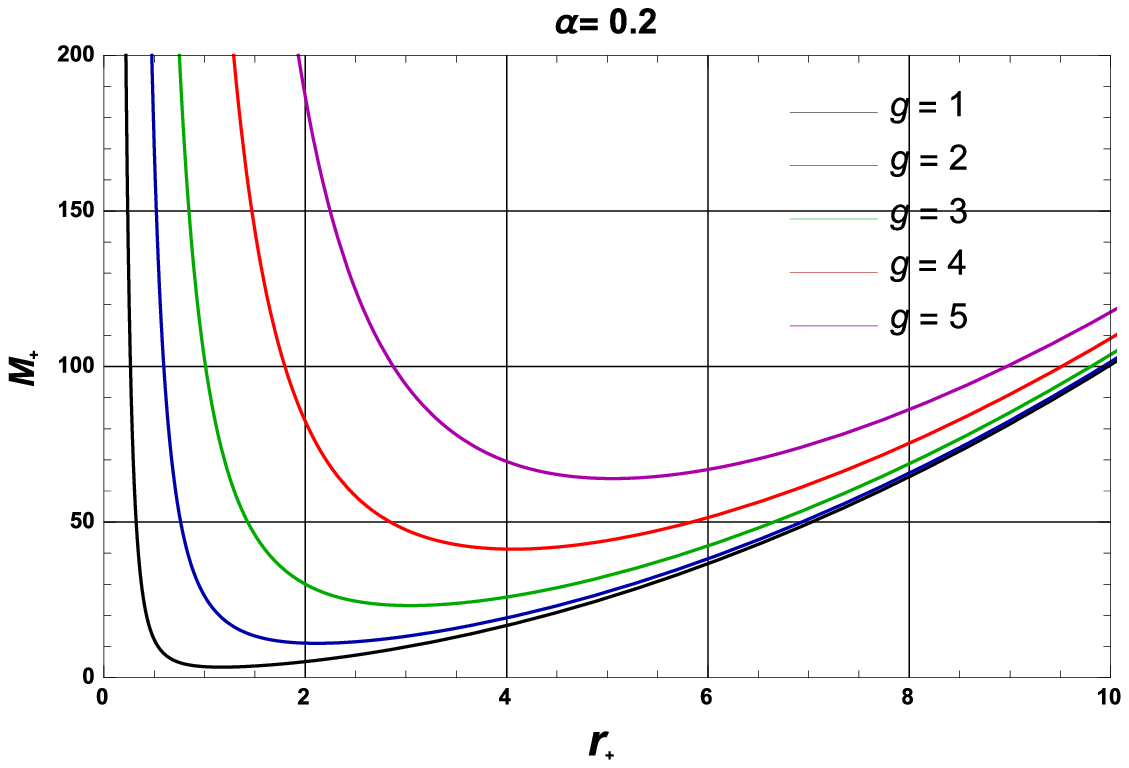}
\end{tabular}
\caption{The plot of mass $M_+$ as the function horizon radius $r_+$ for  different values of parameters $g$ and  $\alpha$.}
\label{rem2}
\end{figure*}
To get black hole remnant we  solve $f'(r_E)=r_E^5-g^3(r_E^2+4\alpha)=0$ for $r_E$ \cite{hamid18} and the results of remnants size, minimum mass and maximum temperature for different value of parameters $g$ and $\alpha$ shown  in Table \ref{tab3}. 

The emitted feature of such a regular black hole can now be simply analyzed by displaying the temporal component of the metric as a function of radius for an extremal EGB-Bardeen black hole with different values of $g$. This has been presented in Fig. \ref{rem1}. This figure exhibits the possibility of having an extremal configuration with one degenerate event horizon at a minimal nonzero mass $M_0$ corresponding to zero temperature. In fact, the condition for having  a degenerate  horizon is that $M = M_0$ which means for $M < M_0$ there are no  horizons (cf. Fig. \ref{rem1} and Fig. \ref{rem2}). In the near extremal region the temperature increases from zero to local maximum $T_{max}$ corresponding to $r_+=r_{max}$. As $r_+$ increase further $T_+$ drops to a local minimum corresponding to $r_+=r_{min}$ and then start increasing linearly.
  \begin{center}
\begin{table}[h]
\begin{center}
\begin{tabular}{l|l r l| r l r l r}
\hline
\hline
\multicolumn{1}{c|}{GB coupling }&\multicolumn{1}{c}{ }&\multicolumn{1}{c}{$\alpha=0.1$  }&\multicolumn{1}{c|}{ \,\,\,\,\,\, }&\multicolumn{1}{c}{ }&\multicolumn{1}{c}{}&\multicolumn{1}{c}{ $\alpha=0.2$ }&\multicolumn{1}{c}{}\,\,\,\,\,\,\\
\hline
\multicolumn{1}{c|}{ { Parameter $(g)$}} & \multicolumn{1}{c}{ $r_C$ } & \multicolumn{1}{c}{ $M_0$}& \multicolumn{1}{c|}{$T_{ma    x}$}&\multicolumn{1}{c}{{$r_C$}}& \multicolumn{1}{c}{$M_0$} &\multicolumn{1}{c}{$T_{max}$}  \\
\hline

\,\, $g=1$\,\,& \,\,1.16\,\, &\,\,  2.97\,\,& \,\,0.055\,\, &\,\,1.18\,\,&\,\,3.37\,\,&\,\,0.050\,\,
\\
\
\,\, $g=2$\,\, & \,\,2.30\,\, &\,\,  10.55\,\,& \,\,0.030\,\,&\,\,2.10\,\,&\,\,11.2\,\,&\,\,0.029\,\,
\\
\
\,\, $g=3$\,\, & \,\,3.43\,\, &\,\,  23.15\,\,& \,\,0.020\,\,&\,\,3.10\,\,&\,\,23.6\,\,&\,\,0.020\,\,
\\
\
\,\, $g=4$\,\, & \,\,4.59\,\, &\,\,  40.70\,\,& \,\,0.015\,\,&\,\,4.06\,\,&\,\,41.2\,\,&\,\,0.015\,\,
\\
\
\,\, $g=5$\,\, & \,\,5.72\,\, &\,\,  63.30\,\,& \,\,0.012\,\,&\,\,5.07\,\,&\,\,63.8\,\,&\,\,0.012\,\,
\\
\hline
 \hline
\end{tabular}
\end{center}
\caption{The critical radius $r_C$, remnant mass $M_0$, and maximum temperature $T_{max}$ for different values of  parameter $g$.}
\label{tab3}
\end{table}
\end{center}
The temperature decreases with increasing value of the critical $r_{C}$ and becomes zero when the two horizons coincide, i.e.,  $T\to 0$ as $r_- \to r_+$. Hence, the $5D$ EGB Bardeen black hole  has a stable remnant with $M=M_0$ and  admits phase transition with divergence of a specific heat  at the critical radius. The black hole cools down with regular double horizon remnant. 

\section{Conclusion}
We find 5D Bardeen-like black holes  as   an exact  solution of EGB gravity minimally coupled to NED and  the famous Boulware-Desser black holes is encompassed as a special case in the absence of NED ($g=0$).   In turn,  we characterized the solution by analyzing horizons which are maximum two, viz. inner Cauchy ($ r_- $) and outer event horizons ($ r_- $). The regularity of EGB-Bardeen black hole spacetime is confirmed by calculating  curvature invariant and they are shown to be well behaved everywhere including at the origin. We have thoroughly analyzed thermodynamics  to analytically compute thermodynamical quantities like the Hawking temperature, entropy, specific heat and free energy associated with  EGB-Bardeen black holes with a focus on the stability of system. It turns out that the heat capacity blows at horizon radius $r_+^C$ which is a double horizon and incidentally  local maxima of the Hawking temperature also occur at $r_+^C$. It is shown that the heat capacity is positive for $r<r_+^C$ suggesting the stability of small black holes against perturbations in the region, and the phase transition exists at $r_+^C$.  While the black hole is unstable for $r>r_+^C$ with negative heat capacity. The global analysis of stability of black holes is also analyzed by calculating free energy $F_+$.  It turns out that the smaller black hole are globally stable with positive heat capacity $C_+>0$ and negative free energy $F_+<0$. Finally, we have also shown that the black hole evaporation result into a  stable black hole remnant with zero temperature $T_+=0$ and positive specific heat $C_+>0$.   

Further,  generalization of such a  regular black hole configuration to Lovelock gravity is an important direction for future which is being under active consideration. In addition, it would be also interesting to generalize this solution to include dS/AdS background. 

\begin{acknowledgements}
  S.G.G. would like to thank  DST INDO-SA bilateral project DST/INT/South Africa/P-06/2016 and also to IUCAA, Pune for the hospitality while this work was being done.
\end{acknowledgements}


\begin{thebibliography}{99}

\bibitem{rp} R. Penrose, Riv. Nuovo Cimento 1, 252 (1969); in General Relativity, an Einstein
	Centenary Volume, edited by S. W. Hawking and W. Israel (Cambridge University
	Press, Cambridge, England, 1979).

	\bibitem{Sakharov:1966}
	A. D.~Sakharov,
	Sov.\ Phys.\ JETP {\bf 22}, 241 (1966).
	
	\bibitem{Gliner:1966}
	E. B.~Gliner,
	Sov.\ Phys.\ JETP {\bf 22}, 378 (1966).
	
\bibitem{Bardeen:1968}	J.~Bardeen, in {\it Proceedings of GR5} (Tiflis, U.S.S.R., 1968).
\bibitem{AGB}E. Ayon-Beato and A. Garcia, Phys. Lett. B {\bf 493}, 149 (2000); Gen.\ Rel.\ Grav.\  {\bf 31}, 629 (1999);  Gen. Rel. Grav. {\bf 37}, 635 (2005).
\bibitem{Ghosh:2015pra} 
S.~G.~Ghosh and M.~Amir,
Eur.\ Phys.\ J.\ C {\bf 75}, 553 (2015).

	
\bibitem{Borde:1994ai}
A.~Borde, 
Phys.\ Rev.\ D {\bf 50}, 3692 (1994); Phys.\ Rev.\ D {\bf 55}, 7615 (1997).
	
\bibitem{AyonBeato:1998ub} E.~Ayon-Beato and A.~Garcia,
	Phys.\ Rev.\ Lett.\  {\bf 80}, 5056 (1998); K. A.~Bronnikov,
	Phys.\ Rev.\ D {\bf 63}, 044005 (2001); S. A.~Hayward,
	Phys.\ Rev.\ Lett.\  {\bf 96}, 031103 (2006);  O. B.~Zaslavskii,
	Phys.\ Rev.\ D {\bf 80}, 064034 (2009);
	J. P. S.~Lemos and V. T.~Zanchin,
	Phys.\ Rev.\ D {\bf 83}, 124005 (2011); I. Dymnikova, Gen. Rel. Grav. {\bf 24}, 235 (1992);  I. Dymnikova, Class. Quant. Grav. {\bf 21}, 4417  (2004); K. A. Bronnikov, Phys. Rev. D {\bf 63},  044005 (2001); S.~Shankaranarayanan and N.~Dadhich, Int.\ J.\ Mod.\ Phys.\ D {\bf 13}, 1095 (2004).
\bibitem{Xiang} L.~Xiang, Y.~Ling and Y.~G.~Shen, Int.\ J.\ Mod.\ Phys.\ D {\bf 22}, 1342016 (2013);
 H.~Culetu,  Int.\ J.\ Theor.\ Phys.\  {\bf 54},  2855 (2015);
 L.~Balart and E.~C.~Vagenas,  Phys.\ Lett.\ B {\bf 730}, 14 (2014);
  L.~Balart and E.~C.~Vagenas,  Phys.\ Rev.\ D {\bf 90}, 124045 (2014);
J.~C.~S.~Neves and A.~Saa,  Phys.\ Lett.\ B {\bf 734}, 44 (2014); 
  D.~V.~Singh, M.~S.~Ali and S.~G.~Ghosh,
Int. J. Mod. Phys. D {\bf 27} 1850108 (2018); D.~V.~Singh and S.~Siwach, arXiv:1909.11529 [hep-th];
  D.~V.~Singh and N.~K.~Singh,
  Ann. Phys.\  {\bf 383}, 600 (2017).
\bibitem{fr1}
S. Fernando, ” Int. J. of Mod. Phys. D {\bf 26}, 1750071 (2017).
\bibitem{Toshmatov} B.~Toshmatov, B.~Ahmedov, A.~Abdujabbarov and Z.~Stuchlik, Phys.\ Rev.\ D {\bf 89}, 104017 (2014);
M.~Amir, F.~Ahmed and S.~G.~Ghosh, Eur.\ Phys.\ J.\ C {\bf 76}, 532 (2016);
A.~Abdujabbarov, M.~Amir, B.~Ahmedov and S.~G.~Ghosh,
Phys.\ Rev.\ D {\bf 93}, 104004 (2016);  C.~Bambi and L.~Modesto,  Phys.\ Lett.\ B {\bf 721}, 329 (2013);
M.~Amir and S.~G.~Ghosh, Phys.\ Rev.\ D {\bf 94}, 024054 (2016);
JHEP {\bf 1507}, 015 (2015); S.~G.~Ghosh, Eur.\ Phys.\ J.\ C {\bf 75}, 532 (2015).

\bibitem{Ghosh} S.~G.~Ghosh and S.~D.~Maharaj, Eur.\ Phys.\ J.\ C {\bf 75}, 7 (2015).
\bibitem{gross1}
D. J. Gross and E. Witten, Nucl. Phys. B {\bf 277}, 1 (1986).
\bibitem{Boulware85}
 D. G. Boulware and S. Deser, Phys. Rev. Lett. {\bf 55}, 2656 (1985).
\bibitem{Wiltshare88}
D. Wiltshire, Phys. Rev. D {\bf 38}, 2445 (1988).
\bibitem{ghosh14} 
S.G.~Ghosh, U.~Papnoi and S.D.~Maharaj,
Phys.\ Rev.\ D {\bf 90},  044068 (2014).
\bibitem{Dadhich:2012cv}
 S.G. Ghosh and S.D. Maharaj, Phys. Rev. D, {\bf 89}, 084027 (2014); S.H. Mazharimousavi and M. Halilsoy, Phys. Lett. B {\bf 681}, 190 (2009);  E. Herscovich and M.G. Richarte, Phys. Lett. B {\bf 689}, 192 (2010);  S.H. Mazharimousavi, O. Gurtug, and M. Halilsoy, Class. Quan. Grav. {\bf 27}, 205022 (2010).

\bibitem{Dadhich:2013bya} 
N.~Dadhich, S.~G.~Ghosh and S.~Jhingan,
Phys.\ Rev.\ D {\bf 88}, 084024 (2013); S.~G.~Ghosh and S.~Jhingan,
Phys.\ Rev.\ D {\bf 82}, 024017 (2010);  S.~Jhingan and S.~G.~Ghosh,
Phys.\ Rev.\ D {\bf 81}, 024010 (2010);
N.~Dadhich, S.~G.~Ghosh and S.~Jhingan,
Phys.\ Lett.\ B {\bf 711}, 196 (2012).
\bibitem{dharm}
S. G. Ghosh, D. V. Singh and S. D. Maharaj, Phys. Rev. D {\bf 97}, 104050 (2018).
\bibitem{jmm} J.M. Maldacena, Adv. Theor. Math. Phys. {\bf 2}, 231 (1998)
\bibitem{kanti} P. Kanti, Black holes at the LHC, in {\it Physics of Black Holes: A Guided Tour, Lecture Notes
in Physics}, vol. 769, ed. by E. Papantonopoulos (Springer, Berlin; New York, 2009), pp. 387–
423.
\bibitem{cvetic} M. Cvetic, H. Lu, D.N. Page, C.N. Pope, J. High Energy Phys. {\bf 07}, 082 (2009).
\bibitem{js} J. J. Schwarz, { Nucl. Phys.} B {\bf 226}, 269 (1983).
\bibitem{vafa} A. Strominger, C. Vafa,  Phys. Lett. B {\bf 379}, 99 (1996).
\bibitem{mayers} R.C. Myers and M.J. Perry, Ann. Phys. (N.Y.) {\bf 172}, 304(1986).
\bibitem{dianyan} X. Dianyan.  Class. Quant. Grav. {\bf 5}, 871 (1988).
\bibitem{Dadhich:2003gw} N.~Dadhich, S.~G.~Ghosh and D.~W.~Deshkar,
Int.\ J.\ Mod.\ Phys.\ A {\bf 20}, 1495 (2005); S.~G.~Ghosh and D.~W.~Deshkar,
Int.\ J.\ Mod.\ Phys.\ D {\bf 12}, 913 (2003);
S.~G.~Ghosh and A.~Banerjee,
Int.\ J.\ Mod.\ Phys.\ D {\bf 12}, 639 (2003);
S.~G.~Ghosh and A.~Beesham,
Phys.\ Rev.\ D {\bf 64}, 124005 (2001); S.~G.~Ghosh and N.~Dadhich,
Phys.\ Rev.\ D {\bf 64}, 047501 (2001);  K. D. Krori, P. Borgohain, and Kanika Das.  J. Math. Phys. {\bf 30}, 2315 (1989).


\bibitem{Ghosh:2006ab} S.~G.~Ghosh and D.~W.~Deshkar,
Astrophys.\ Space Sci.\  {\bf 310}, 111 (2007);
S.~G.~Ghosh and D.~W.~Deshkar,
Phys.\ Rev.\ D {\bf 77}, 047504 (2008);
S.~G.~Ghosh, S.~B.~Sarwe and R.~V.~Saraykar,
Phys.\ Rev.\ D {\bf 66}, 084006 (2002);
N.~Dadhich and S.~G.~Ghosh,
Phys.\ Lett.\ B {\bf 518}, 1 (2001);
S.~G.~Ghosh and R.~V.~Saraykar,
Phys.\ Rev.\ D {\bf 62}, 107502 (2000);
S.~G.~Ghosh, D.~W.~Deshkar and N.~N.~Saste,
Int.\ J.\ Mod.\ Phys.\ D {\bf 16}, 53 (2007).


\bibitem{sgg03} S.G. Ghosh and D.W. Deshkar. 
Int. J. Mod. Phys. D {\bf 12} 913 (2003); S.G. Ghosh and A. Banerjee. Int. J. Mod. Phys. D {\bf 12}
639 (2003);  S.G. Ghosh and N. Dadhich.  Phys. Rev. D {\bf 65} 127502 (2002).



\bibitem{egb}
C. Lanczos, Ann. Math. {\bf 39}, 842 (1938).
\bibitem{ll}
D. Lovelock, J. Math. Phys.  {\bf 12}, 498 (1971).

\bibitem{gross}
D. J. Gross and J. H Sloan, Nucl. Phys. B {\bf 291} 41 (1987); M. C. Bento and O. Bertolami, Phys. Lett. B {\bf 368} 198 (1996).
\bibitem{ghosh8}
S. G. Ghosh and D. W. Deshkar  Phys. Rev. D {\bf 77}, 04750.
\bibitem{Antoniou:2017hxj} 
  G.~Antoniou, A.~Bakopoulos and P.~Kanti,
  Phys.\ Rev.\ D {\bf 97}, no. 8, 084037 (2018); J.~L.~Blázquez-Salcedo, D.~D.~Doneva, J.~Kunz and S.~S.~Yazadjiev,
  arXiv:1805.05755 [gr-qc].
\bibitem{Hendi:2017phi}
  S.~H.~Hendi, N.~Riazi, S.~Panahiyan and B.~Eslam Panah,
  arXiv:1710.01818 [gr-qc];
  S.~Panahiyan, S.~H.~Hendi and N.~Riazi,
  arXiv:1802.00309 [gr-qc].
\bibitem{Ghosh:2016ddh}
  S.~G.~Ghosh, M.~Amir and S.~D.~Maharaj,
  Eur.\ Phys.\ J.\ C {\bf 77},  530 (2017).

\bibitem{rizzo06}
T.  G. Rizzo JHEP {\bf 09}, 021 (2006).
\bibitem{sabir}
Md S. Ali, S. G. Ghosh,  Phys. Rev. D {\bf 98} 084025 (2018).



\bibitem{myers88}
R. C. Myers and J. Z. Simon, Phys. Rev. {\bf D 38} 2434 (1988). 

\bibitem{hp}S. Hawking and D. Page, Commun. Math. Phys. {\bf 87}, 577 (1983).

\bibitem{cai02}
 R. G. Cai, Phys. Rev. D {\bf 65}, 084014 (2002);
R. C. Myers and J. Z. Simon, Phys. Rev. D {\bf 38}, 2434 (1987).

\bibitem{Maluf:2018lyu}
  R.~V.~Maluf and J.~C.~S.~Neves,
  Phys.\ Rev.\ D {\bf 97}, 104015 (2018).
\bibitem{davis77}
P. Davis, Proc. R. Soc. A {\bf 353}, 499 (1977).

\bibitem{Herscovich}
E. Herscovich and M. G. Richarte, Phys. Lett. B {\bf 689}, 192–200 (2010). 
\bibitem{jp}J. Preskill, Do black hole destroy information, arXiv:9209058 [hep-th].
\bibitem{jh}  J.H. MacGibbon, Nature {\bf 329} 308 (1987).
\bibitem{hamid18}
S. Hamid Mehdipour and M. H. Ahmadi, Nuc. Phys. {\bf B 926} 49 (2018);
I. G. Dymnikova, Int. Journal of Mod. Phys. D, {\bf 5}  529 (1996);
I.  G. Dymnikova and M. Korpusik, Phys. Lett. B {\bf 685}, 12–18 (2010). 
\bibitem{ads}
A. Kumar, D. V. Singh and S. G. Ghosh, Eur. Phys. J. C 79, 275 (2019)..
\end{thebibliography}
\end{document}